\documentstyle[prb,twocolumn,aps]{revtex}

\input{epsf}

\begin{document}
\tolerance 10000

\draft

\title{Evidence for Quasiparticle Decay in Photoemission from
      Underdoped Cuprates}

\author{R. B. Laughlin}


\address{
Department of Physics,
Stanford University, Stanford, California 94305 USA}

\twocolumn[
\date{Received}
\maketitle
\widetext

\vspace*{-1.0truecm}

\begin{abstract}
\begin{center}
\parbox{14cm}{I argue that the ``gap'' recently observed at the
Brillouin zone face of cuprate superconductors in photoemission by
Marshall et al [Phys. Rev. Lett. {\bf 76}, 4841 (1996)] and Ding et
al [Nature {\bf 382}, 54 (1996)] is evidence for the decay of the
injected hole into a spinon-holon pair.}
\end{center}
\end{abstract}

\pacs{
\hspace{1.9cm}
}
]

\narrowtext

One of the most interesting developments in cuprate superconductivity
is the the recent observation by Marshall et al \cite{marshall} and
Ding et al \cite{ding} of a pseudogap in the electron spectral
function near the Brillouin zone face that persists above the
superconducting transition temperature and grows in magnitude as doping
is reduced.  This feature, which is also seen in optical
conductivity \cite{homes} and is almost certainly the ``spin gap''
effect seen in magnetic resonance \cite{resonance}, has the momentum
dependence expected of a simple d-wave superconductor but a size,
doping dependence, and breadth that do not, particularly at low
dopings.

The purpose of this paper is to propose that these experiments may
constitute long-sought evidence that spinons and holons, the
soliton-like particles known from studies of 1-dimensional
antiferromagnets \cite{anderson}, actually exist in these materials.
The reason is that there is no other simple explanation of the
experiments that is not also contrived. For example, the evolution
of the feature out of the d-wave gap with underdoping has led to
speculation that it is the dissociation of a ``pre-formed'' Cooper pair,
this being a specific realization of the quite sensible ideas of
Kivelson and Emery \cite{kivelson}.  However, the attractive force
required to accomplish such pairing would be outrageously large, no
such effect has ever been observed in a conventional metal, and the
effect persists to extreme underdoping where the material is an
insulator.  Similarly, the practice of modeling the system as a spin
density wave does not work in situations lacking long-range order,
requires delicate adjustments of the distant-neighbor hopping integrals
to account for the observed isotropy of the quasiparticle dispersion
relation, and does not account at all for the enormous width of the
quasiparticle peak at the zone face. The last two remarks apply broadly
to existing work on the $t-t'-J$ model as well \cite{ttj}.  The
discussion I shall present deliberately avoids sophisticated mathematics
and argues directly from the experiments shown in Figs. 1 and 2 and the
consistency of these with Eqs. (1) - (4).  A formalism-free approach
is essential because the objective is not to make a theory of high-Tc
superconductivity - a delicate question of symmetry breaking - to
promote a model or to report calculations, but rather to establish that
spinons and holons are real.

\begin{figure}
\epsfbox{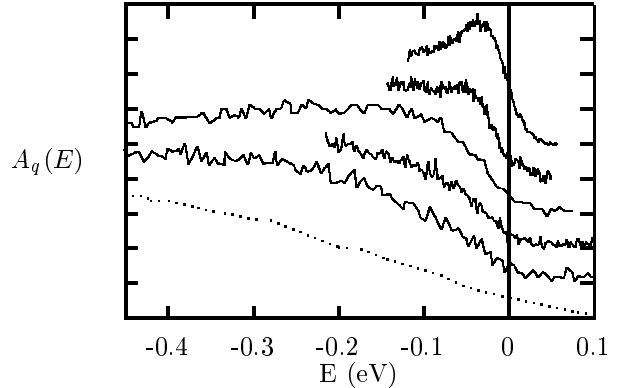}

\caption{Solid: Photoemission energy distribution curves near the
X-point of BSCCO at various dopings as described in the text.
Dashed: $X'$ curve of the magnetic insulator Sr$_2$CuO$_2$Cl$_2$
taken from Fig. 2 with the zero of energy shifted by 0.7 eV to account
for the chemical potential difference.}

\end{figure}

The spinon and holon I have proposed to be present in the cuprates have
the dispersion relations

\begin{equation}
E_{k}^{spinon} = 1.6 J \sqrt{ \cos^2 (k_{x}b) + \cos^2 (k_{y}b) }
\; \; \; ,
\end{equation}

\begin{equation}
E_{k}^{holon} = \pm 2t \sqrt{ \cos^2 (k_{x}b) + \cos^2 (k_{y}b) }
\; \; \; ,
\end{equation}

\noindent
where t = 0.5 eV and J = 0.125 eV are the bandwidth and magnetic
exchange parameters of a magnetic Hamiltonian such as the t-J model
and $b = 4$ \AA is the bond length.
I wish to be somewhat vague about the specifics of the Hamiltonian
because it is not known whether any such model describes the cuprates
in detail.  Fortunately Eqs. (1) and (2), unlike questions of order,
are insensitive to subtleties.  The values of the parameters are
important. t is a tight-binding fit to the bare Hartree-Fock-Slater
band structure \cite{mattheiss} and is a number characterizing charge
transport. J is a Heisenberg fit to the 2-magnon raman scattering
\cite{raman} and neutron scattering \cite{hayden} experiments performed
on the insulator and is a number characterizing the magnetism.  Both
parameters should be considered known and not adjusted later to fit
other experiments.

The ``new'' development motivating this paper is the discovery of the
spin gap in underdoped superconductors.  I have recently written a
series of papers arguing that spinons and holons may be seen indirectly
in numerical studies of the t-J model \cite{dagotto} through subtle
inconsistencies of sum rules as t and J are varied \cite{rbl}.  A more
direct observation has not been possible - until now - because the
violent attraction of these particles for each other so distorts the
commonly calculated spectra that their shape becomes mostly a measure
of the interaction \cite{rbl}. These experiments, however, imply that
the interaction is substantially weaker in the cuprates than it is in
the t-J model at low doping. There are many potential causes of this -
doping, elevation of the temperature, or modification to the
Hamiltonian, for example - and distinguishing among these is quite
beyond our means at present. So we must defer the question
of cause for now, or more precisely restrict ourselves to versions of
the question that have experimental answers.  For example, I currently
favor the theory that the finite temperature required to prevent the
sample from charging is the cause of this weakness in the insulators.
The only reasonable test of this is to repeat the experiment cold and
see if the t-J results materialize. But regardless of the cause, the
weakened interaction is the key difference between the experiments
and the t-J studies at low doping.

Let me begin by making a connection between the insulators and metals.
In the inset of Fig. 1 I show a compilation of energy distribution
curves taken by two different experimental groups near the X-point of
BSCCO for various dopings.  The samples were made in different
laboratories and have slightly different stoichoimetries. The short
curves are, top to bottom, the $T_{c}$ = 87 K, 83 K, and 10 K data
taken from Fig. 1 of Ding et al \cite{ding}, which correspond to
samples of Bi$_{2}$Sr$_{2}$CaCu$_{2}$O$_{8 + \delta}$ measured at the
$(\pi , 0) - (\pi ,\pi)$ fermi surface crossing. The long curves
are, top to bottom, the Tc = 65 K $(\pi ,0)$ curve of Fig. 2 of
Marshall et al \cite{marshall}, which corresponds to a
Bi$_{2}$Sr$_{2}$Ca$_{1-x}$Dy$_{x}$Cu$_{2}$O$_{8+\delta}$
sample at 10\% Dy doping, and unpublished \cite{thesis} data for an
insulating sample of this material with 35\% Dy doping.  When they
are plotted on the same graph in this way it becomes obvious that the
data on the underdoped superconductors {\it interpolate} between the
behavior found at optimal doping, which is roughly consistent with
conventional metal physics, and the behavior reported by Wells et al
\cite{wells} for the magnetic insulator Sr$_{2}$CuO$_{2}$Cl$_{2}$
reproduced in Fig. 2.  This fact has two major implications. The first
is that the strange behavior of the insulator near the X-point is not
an artifact of the particular material, but is generic to the cuprates
and therefore worth understanding.  This was not clear when it was
first discovered. The second is that it is the same effect as the spin
gap. This is an extremely strong statement but it is clearly true, for
otherwise we would need to invent two independent mechanisms for
producing ``d-wave'' gapping in this problem and explain why one of
them continuously evolved into the other with doping.  The study of
the insulator and the study of the spin gap are the same thing.

\begin{figure}
\epsfbox{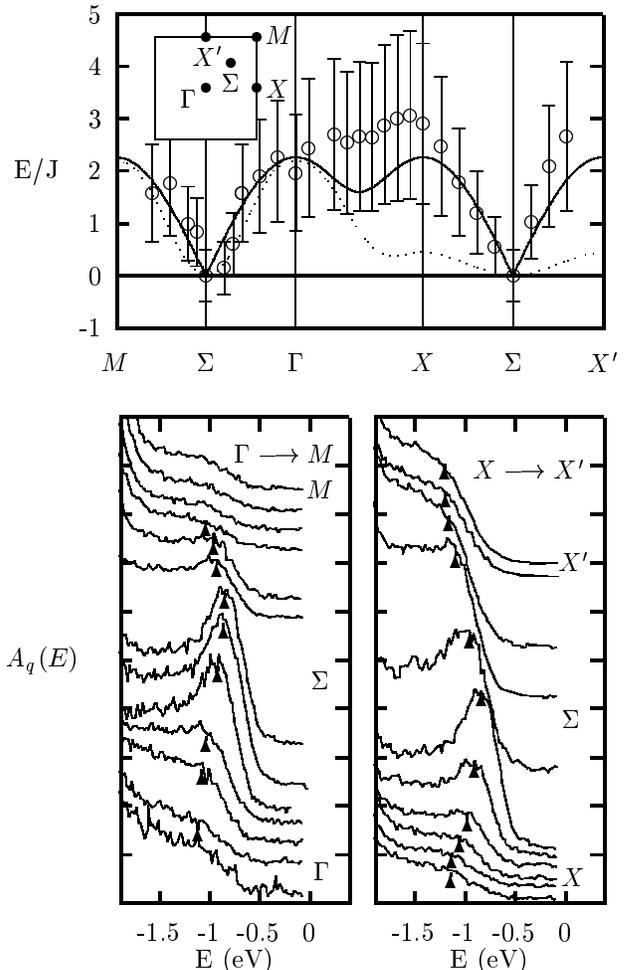}

\caption{Top: Comparison of ``quasiparticle'' dispersion relation found
in insulating Sr$_{2}$CuO$_{2}$Cl$_{2}$ by Wells et al with the
prediction of Eq. (1).  The error bars represent my estimate of the
width of the peak and not the accuracy of the measurement.
The dashed curve is the dispersion relation found by
numerical t-J studies.  Bottom: Photoemission energy distribution
curves from which this dispersion relation was inferred. The arrows
correspond to the open circles in the top panel.}

\end{figure}

Let us now consider the question of quasiparticle integrity. It may be
seen in Fig. 2 that the energy distribution curves of
Sr$_{2}$CuO$_{2}$Cl$_{2}$ show a peak that disperses with momentum,
has its lowest energy at $\Sigma$, and broadens substantially away from
this minimum.  Along the line $\Gamma \rightarrow M$ the data are
essentially indistinguishable from those of the superconductors at any
doping, not only in shape but in scale, so nothing is lost by plotting
only the insulator data.  It may also be seen that the spectra at the
extremal points $\Gamma$, $M$, and $X$ are so broad that they are more
properly characterized as a continuum with the hint of a knee or edge
about 0.2 eV above the fermi energy.  This edge was interpreted by
Wells et al \cite{wells} as the quasiparticle peak at this momentum
and plotted as a point in their quasiparticle dispersion relation, also
reproduced in Fig. 2.  While this is a reasonable thing to do if the
quasiparticle is assumed to exist, it is very {\it un}reasonable
otherwise, for the ostensible lifetime broadening at these momenta
is greater than the entire dispersion across the zone.  We know this
broadening to be intrinsic and not caused by surface disorder because
the same sample shows a well-defined peak near $\Sigma$ and strong
angle dependence of this peak.  Also, {\it all} the cuprates show such
broadening at $\Gamma$ and $M$.  So these data actually imply that the
quasiparticle has no integrity at these momenta at all, and does not,
in fact, exist.

Let us now consider the energy scale of the quasiparticle.  It is now
well established that the quasiparticle bandwidth in all the cuprates
is about 0.3 eV regardless of crystal structure or doping level, a
rather astonishing fact in light of their different transport and
optical properties.  It may be seen in Fig. 2, for example, that the
quasiparticle energy at $\Gamma$, $M$, and $X$, insofar as it is
defined, is 0.3 eV higher than that at $\Sigma$.  This energy scale
is an important clue to the nature of the microscopic physics because
it is so peculiar. Conventional metals and semiconductors have
bandwidths ten times larger than this because their energy scale is
set by the matrix element for electrons to hop between adjacent sites
- typically 1 or 2 eV.  This is why the cuprate bandwidth is 2-3
times smaller than that predicted by conventional Hartree-Fock-Slater
band structure calculations \cite{mattheiss} and well outside their
expected error bar. Nor is it reasonable to ascribe this energy to
phonons.  The cuprates are ionic and thus have large electron-phonon
couplings, but not larger than those in alkalai halides, where the
effect of phonons is either to enhance the band mass slightly, as
occurs in the conduction band, or to enhance it by many orders of
magnitude through the small polaron effect, as occurs in the valence
band.  Indeed the only energy in the problem the right size to account
naturally for this bandwidth is the magnetic exchange parameter J.
One of the strangest and most consistent findings of the numerical work
on the t-J model \cite{poilblanc} at low doping has been that the
quasiparticle bandwidth is 2.2 J  regardless of the value of t
\cite{rbl}.  Since the bandwidth does not require high resolution to
compute and is known to be relatively insensitive to other parameters
such as $t'$, it must be considered a firm prediction of these
calculations that the bandwidth should be about 0.3 eV.  Thus the
agreement between the prediction of the model and experiment suggests
that the t-J model has some relevance to the problem and that
the quasiparticle bandwidth is set by J.

Let us finally consider the question of isotropy.  It may be seen from
Fig. 2 that the energy scale of the quasiparticle at $\Gamma$, $M$, and
$X$ is the same, and that the dispersion near $\Sigma$, where the peak
is sharpest, is isotropic.  This isotropy does {\it not} agree with the
t-J studies at low doping, which match experiment in the $\Gamma
\rightarrow M$ direction but show no dispersion at all $X \rightarrow
X'$ direction.  This disparity has led a number of theorists to add
other parameters, typically a second-neighbor hopping integral $t'$,
to the Hamiltonian and adjust its value to make the quasiparticle
dispersion in an approximate calculation match experiment everywhere.
In addition to failing to account for the quasiparticle width, this
line of reasoning has the obvious flaw of ascribing the isotropy to a
coincidence of the parameters J and $t'$, notwithstanding the
strangeness of the energy scale.  While such a coincidence is
conceivable, it is far more reasonable to conclude that the dispersion
in the $X \rightarrow X'$ direction is regulated by the same parameter
regulating the $\Gamma \rightarrow M$ dispersion, namely J, and that
the failure of the t-J calculations to find this effect is a subtle
problem related to their failure to find the correct quasiparticle
width at X.

\begin{figure}
\epsfbox{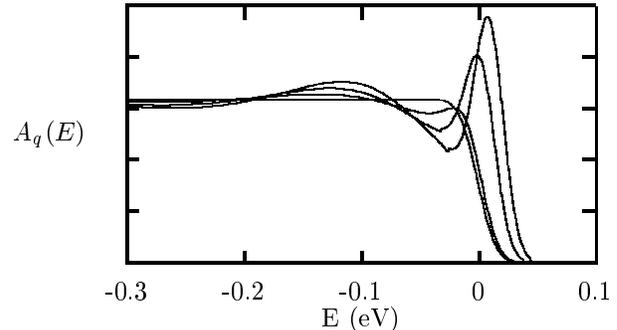}

\caption{Theoretical spectral function defined by Eq. (3) for $V_0 =$
0.00 eV, 0.01 eV, 0.02 eV, and 0.03 eV.  The spectrum has been
convolved with a gaussian of width 0.012 eV.}

\end{figure}

I now turn to my interpretation of these experiments, namely that the
injected hole is decaying into a spinon-holon pair.  The disparity
between t and J in our problem is such that decay is expected to result
in a continuum several electron volts wide and an {\it edge} tracing out
the spinon dispersion relation, the minimum-energy decay being when the
spinon carries away all the momentum.  If there is in addition a weak
attractive force between the spinon and holon, structure develops
at this edge.  In Fig.3 I plot the spectrum

\begin{equation}
A(E) = \sum_n | \psi_n (0) |^2 \; \delta ( E + E_n )
\; \; \; ,
\end{equation}

\noindent
where

\begin{equation}
{\cal H} \psi_n (r) = E_n \; \psi_n (r)
\end{equation}

\begin{equation}
{\cal H} = - \frac{\hbar^2}{2m^*} \nabla^2 - V_0 \; \theta(r_0 - r)
\; \; \; ,
\end{equation}

\noindent
which is a model 2-body Greens' function matrix element for the two
particles to coincide in space in the limit that the spinon bandwidth
is zero.  The parameter $m^* = \hbar^2 / (\sqrt{2} t b^2 )$ is the holon
band effective mass computed from Eq. (2) or 0.7 electron masses, the
square-well radius is $r_0 = 6 b$ or 24 \AA, and the depth is
$0.0 eV \leq V_0 \leq 0.03 eV$.  The ``quasiparticle'' peak in this
spectrum is a bound state of the spinon and holon which is small
because the wavefunction in question is physically large.  The
features in the spectrum at higher energy are scattering resonances;
note that these occur at an energy scale 10 times
larger than the size of $V_0$, which is itself of order the
superconducting $T_c$. In Fig. 2 I compare the ``quasiparticle''
dispersion relation inferred by Wells et al \cite{wells} with the
spinon dispersion relation of Eq. (1).  The agreement between the two
is obviously excellent, including the overall energy scale, which is
not adjustable, the equivalence of the energies at $\Gamma$, $M$, and
$X$, and the isotropy near $\Sigma$. Thus except for the peaking of
the spectrum near $\Sigma$, which I ascribe to dependence of the
attractive force on the center-of-mass momentum, this experiment,
and the spin gap experiments like it, are plausibly understood as the
broad continuum expected from quasiparticle decay with a threshold at
the energy of the bare spinon.

As Eqs. (1) and (2) are essential to the argument, let me now explain
their origin.  These are the dispersion relations for the spinon and
holon found by a number of us in the early days of high-$T_{c}$.  They
are alternately described in papers of that time as behavior of an
electron in a d-wave superconductor \cite{zanderson}, the behavior of
an electron in a magnetic field of flux $\pi$ per plaquette
\cite{affleck}, and the behavior of a doubled Dirac fermion on a
square lattice \cite{double}. It later came to be understood that none
of these had its literal physical meaning and all were mathematically
equivalent \cite{kotliar}. The spinon and holon are actually fractional
particles analogous to the charge carriers in the fractional quantum
hall effect. The d-wave or flux ``order'' is simply the price one pays
to describe such an object in conventional particle language.  It is
fictitious and disappears when the spinon and holon are written
down as actual spin wavefunctions \cite{zou}. However, these are so
complex that it is usually more convenient to work with an overcomplete
basis and fictitious order.  This is the underlying reason why
descriptions of the antiferromagnet based on these particles are always
gauge theories. The factors of 2t and 1.6 J are fits made by me in
reconciling these equations with known properties of the t-J model.
They have proper justifications within the context of the gauge theory,
but this is less important than the fact that they were published
before the experiments were performed.

Let me now show in a crude way how an attractive force between two
such particles can account for the behavior found in the numerical
work on the t-J model.  A highly detailed explanation is
undesirable here because it would amount to model building and
add unnecessary complexity.  We consider two particles moving on a
square lattice and described by the Hamiltonian

\begin{equation}
{\cal H} = \sum_{<j,k>} V_{jk} \biggl\{ a_{j}^{\dag} a_{k} +
b_{j}^{\dag} b_{k} \biggr\}
- V_{0} \sum_{j} a_{j}^{\dag} b_{j}^{\dag} b_{j} a_{j} \; \; \; ,
\end{equation}

\noindent
where $< \! j,k \!>$ denotes the set of near-neighbor pairs, with each
pair counted twice to maintain hermiticity, and $V_{jk}$ is taken to
be $-V$ for the horizontal bonds in even rows and $+V$ for all
remaining bonds.  When the attractive interaction $V_{0}$ is turned
off the particles are free and are described by the dispersion
relation $E_{k} = \pm 2V\sqrt{ \cos^2 (k_{x}) + \cos^2 (k_{y})}$, which
has the functional form of Eqs. (1) and (2).  When the attraction is
very large, on the other hand, the particles form a bound state, the
dispersion relation of which is $E_{k} = 4 V^{2}/V_{0} [ \cos (k_{x})
+ \cos (k_{y}) ]$, the functional form found in the t-J studies.  The
difference between these two behaviors is due to the bound state's being
a whole particle rather than a fractional one, and thus blind to the
fictitious magnetic field.  The issues of dynamical scale of the
bound state, the presence of two holon branches but only one spinon
branch, and the detailed nature of the attractive force have all
been discussed at length in previous papers \cite{rbl}, but they are
not conceptually important.  The new and important observation is that
the functional form of conventional bands is restored by gauge
invariance when spinons and holons bind tightly, regardless of details.

The issue of whether spinons and holons are real is centrally important
to high-T${c}$ superconductivity.  It must be squarely faced if there is
to be a meaningful discussion of theories based on spin-charge
separation.

I wish to thank Z.-X. Shen, and M. Norman for providing
access to these data and for numerous helpful discussions. This work
was supported by the NSF under grant No. DMR-9421888.

\end{document}